\documentclass[prl,aps,twocolumn,superscriptaddress,floatfix,citeautoscript]{revtex4}
\usepackage{graphicx,rotating,subfigure,amsmath,amsfonts,amssymb,delarray}
\usepackage{dsfont}
\usepackage[T1]{fontenc}

\newcommand{\e}{\text{e}}
\newcommand{\im}{\text{i}}

\def\12{\frac{1}{2}}

\newcommand{\be}{\begin{equation}}
\newcommand{\ee}{\end{equation}}
\newcommand{\bea}{\begin{eqnarray}}
\newcommand{\eea}{\end{eqnarray}}

\renewcommand\Re{\operatorname{Re}}
\renewcommand\Im{\operatorname{Im}}
\DeclareMathOperator\Tr{Tr}

\begin{document}
\bibliographystyle{apsrev}
\title{Closed and Open System Dynamics in a Fermionic Chain with a Microscopically Specified Bath: Relaxation and Thermalization}

\author{Nicholas Sedlmayr}
\affiliation{Department of Physics and Research Center OPTIMAS,
  Technical University Kaiserslautern,
  D-67663 Kaiserslautern, Germany}
\author{Jie Ren}
\affiliation{Department of Physics and Research Center OPTIMAS,
  Technical University Kaiserslautern,
  D-67663 Kaiserslautern, Germany}
\affiliation{Department of Physics and Jiangsu Laboratory of Advanced
Functional Material, Changshu Institute of Technology, Changshu 215500, China}
\author{Florian Gebhard}
\affiliation{Department of Physics, Philipps-Universit\"at Marburg,
35032 Marburg, Germany}
\author{Jesko Sirker}
\affiliation{Department of Physics and Research Center OPTIMAS,
  Technical University Kaiserslautern,
  D-67663 Kaiserslautern, Germany}
\date{\today}

\begin{abstract}
  We study thermalization in a one-dimensional quantum system
  consisting of a noninteracting fermionic chain with each site of
  the chain coupled to an additional bath site. Using a density matrix
  renormalization group algorithm we investigate the time evolution of
  observables in the chain after a quantum quench.  For low densities
  we show that the intermediate time dynamics can be quantitatively
  described by a system of coupled equations of motion.
  For higher densities our numerical results show a prethermalization
  for local observables at intermediate times and a full
  thermalization to the grand canonical ensemble at long times. For
  the case of a weak bath-chain coupling we find, in particular, a
  Fermi momentum distribution in the chain in equilibrium in spite of
  the seemingly oversimplified bath in our model.
\end{abstract}


\maketitle

\paragraph{Introduction.}
The time evolution of classical and quantum systems is deterministic.
If a system in the thermodynamic limit reaches thermal equilibrium at
long times, we expect, however, that its physical properties will be
determined by only a few parameters such as the temperature, chemical
potential, and pressure.  This thermalization process is often studied
in two different settings: (a) The system is in contact with a thermal bath,
i.e., a large reservoir of thermal energy. The key
assumptions commonly used in this setting are a weak coupling between
the bath and system and Markovian dynamics, i.e., a very short correlation
time in the bath.  In this case the microscopic details of the bath
become unimportant~\cite{BreuerPetruccione,Weiss,Carmichael}%
; for a simple example of classical thermalization,
see Ref.~\cite{gebhardmuenster}.
(b)~The
system is closed, with particles being able to exchange energy and
momentum among each other, so that the closed system can explore phase
space, constrained only by the conservation laws such as total energy and
particle number. An important difference between the two scenarios is
that in the first case temperature, chemical potential, and pressure
are parameters determined externally by the bath. In the latter case,
on the other hand, these parameters are Lagrange multipliers fixing
the values of the conserved quantities
\cite{LandauLifshitz3,RigolDunjko}.

In this letter we want to study these two settings simultaneously
using a model which can be either viewed as a closed quantum system or
as a chain coupled to a simple bath.
Thermalization, in both cases, requires: (I) Observables become time
independent and all currents vanish ({\it equilibration}); (II) Time
averages can be replaced by statistical averages over ensembles with a
restricted number of intensive parameters \footnote{Generically
  this number is finite. Here we want to also include the case of
  integrable models in one dimension where this number has to increase
  linearly with system size.}, and are independent of
initial conditions ({\it ergodicity}) \cite{vonNeumann}.  The rather old but
fundamental problem of nonequilibrium dynamics and thermalization in
closed quantum systems has been put again into focus by experiments on
cold quantum gases which are very well isolated from their
surroundings
\cite{TrotzkyChen,KinoshitaWenger,HofferberthLesanovsky,StrohmaierGreif},
as well as by the development of new numerical techniques to study
dynamics in many-body systems
\cite{DaleyKollath,WhiteFeiguin,SirkerKluemperDTMRG,VidalTEBD1,VidalTEBD2,EnssSirker,AndersSchiller,Kehrein}.
This has led to numerous simulations of nonequilibrium dynamics in
closed quantum models where the question of whether or not thermalization
occurs has not always been easy to answer due to the finite numerical
simulation time
\cite{RigolMuramatsu,KollathLauchli,ManmanaWessel,BiroliKollath}.

\paragraph{Closed quantum systems.}
The time evolution of an initial state $|\Psi_0\rangle\equiv
|\Psi(t=0)\rangle$ is unitary and given by the Schr\"odinger equation.
Therefore $|\Psi(t)\rangle$ remains a pure state for all times $t$.
Since ensemble averages describe mixed states such a description
cannot apply to a finite closed quantum system as a whole. Only a
subsystem can be in or close to a thermal state with the rest of the
system acting as an effective bath. Furthermore, contrary to a
classical system, every quantum system has exponentially many
conserved quantities, e.g.~the projection operators
$P_n=|E_n\rangle\langle E_n|$ onto the eigenstates of a system with
a discrete spectrum, $|E_n\rangle$ \cite{RigolDunjko}. However, it is
usually assumed that only the local conserved quantities are of
relevance for thermalization.  A {\it local} conserved quantity can be
represented for a lattice system as $Q_m= \sum_j q^m_j$ where $q^m_j$
is a density operator acting on lattice sites $j,j+1,\cdots,j+m$ with
$m$ finite.
Here we want to concentrate on the case of generic one-dimensional
quantum systems with a small number of local conservation laws,
i.e.~the total energy and particle number.  Thermalization in
closed integrable models, where the number of local conservation laws
increases linearly with the system size \cite{HubbardBook,SirkerLL}, has
been investigated with the help of numerical simulations in recent
times as well \cite{RigolDunjkoPRL,Rigol,RigolPRA,SantosRigol}.

We consider the nonequilibrium dynamics ensuing after
preparing the system in a pure state $|\Psi_0\rangle$ which is not an
eigenstate of the Hamiltonian. Using a Lehmann representation we can
write $|\Psi_0\rangle =\sum_n c_n |E_n\rangle$ where $|E_n\rangle$ are
the eigenstates of the Hamiltonian $H$ the system evolves under.
Furthermore, we restrict ourselves to {\it typical states}
with a macroscopic number $c_n\neq 0$ \footnote{If this condition is
  not fulfilled we cannot expect, in general, that time averages
  become independent of the microscopic details of the initial
  state.}.  We can now easily calculate the long-time mean
\begin{eqnarray}
\label{mean}
\bar{O} &=& \lim_{\tau\to\infty}\sum_{n,m} \frac{1}{\tau}\int_0^\tau \textrm{d}t\, \e^{\textrm{i}(E_m-E_n)t} c_n^* c_m\langle E_m|O|E_n\rangle \nonumber \\
&=& \sum_n |c_n|^2 O_{nn}
\end{eqnarray}
of an observable $O$, where we have set $\hbar=1$. The second line of
Eq.~(\ref{mean}) is often called the {\it diagonal ensemble}. Here we have
assumed that the system is generic, i.e., that degeneracies play no
role. If the observable becomes stationary at long times its
value $O_\infty =\lim_{t\to\infty}\langle\psi(t)| O|\psi(t)\rangle$
has to be equal to the long-time mean, $O_\infty\equiv \bar{O}$. Note
that this is only possible in the thermodynamic limit. Otherwise
observables show revivals on time scales of the order of the system
size. Taking the thermodynamic limit is thus essential; a finite
system can never thermalize.

If a subsystem of an infinite system containing the observable $O$
equilibrates and the value $O_\infty$ does not depend on details of
the initial state, then the remaining open question is which ensemble
describes the equilibrated system. If we have two statistically
independent subsystems $A$ and $B$, the density matrix $\rho$ of
the whole system is given by $\rho=\rho_A\otimes\rho_B$, and thus
$\ln\rho=\ln\rho_A \oplus \ln\rho_B$.  Second, the density matrix
itself should become time independent once the system has equilibrated
and the von Neumann equation implies $\dot\rho
=-\textrm{i}[H,\rho]=0$. Thus the general density matrix under
consideration has to be of the form
$\rho=\exp(-\sum_n\lambda_n\mathcal{Q}_n)/Z$ where $\mathcal{Q}_n$ are
the conserved quantities of the system \cite{LandauLifshitz3}. The
partition function $Z$ is a normalization factor such that
$\Tr\rho=1$. We stress again that the intensive parameters $\lambda_n$
are not given externally but rather are Lagrange multipliers
determined by the set of equations
\begin{equation}
\label{Lagrange}
\langle\Psi_0|\mathcal{Q}_n|\Psi_0\rangle = \Tr\left\{\mathcal{Q}_n\rho\right\}.
\end{equation}
If we include all projection operators into our density matrix,
$\mathcal{Q}_n=P_n$, it follows immediately from Eq.~(\ref{Lagrange})
that $\langle O\rangle_\rho\equiv \Tr\{\rho O\}$ is identical to the
diagonal ensemble as given in Eq.~(\ref{mean}) \cite{RigolDunjko}.
Having to use infinitely many Lagrange multipliers is expected because
$|\Psi(t)\rangle$ is always a pure state and the system as a whole
therefore does not thermalize, because it does not fulfill condition
(II).

In this Letter we focus on the generic situation where we split our
system $S=A\cup B$ into a bath $B$ and a subsystem $A$ and consider
observables acting only on subsystem $A$. We concentrate on the
following questions: How does a subsystem $A$ without intrinsic
relaxation processes equilibrate when coupled to a strongly correlated
but simple and possibly non-Markovian bath $B$? Which statistical
ensemble gives the expectation values of observables in $A$ in the
equilibrated state?

\paragraph{Model Hamiltonian.}
To investigate some aspects of the questions raised above we consider
a simple model system with Hamiltonian \cite{SirkerGebhard}
\begin{eqnarray}
\label{Model}
H&=& -J\sum_{j=1}^{L-1} \left \{ c_j^\dagger c_{j+1} +h.c.\right\}+ \gamma\sum_{j=1}^L\left\{ s^\dagger_j c_j + \textrm{H.c.}\right\} \nonumber\\
&&+V_s\sum_{j=1}^{L-1}\left(s_j^\dagger s_j-1/2\right) \left(s^\dagger_{j+1}s_{j+1}-1/2\right)\;.
\end{eqnarray}
The first term describes a chain of free spinless fermions with
hopping amplitude $J$ and is the subsystem $A$ we study the
thermalization of. The `bath' $B$ consists of extra sites, coupled to
the chain sites via a hybridization $\gamma$ (second term), and we also
include a nearest-neighbor interaction $V_s$ between the bath sites
(third term).

As initial states for the time-evolution with the Hamiltonian
(\ref{Model}) we will consider, on the one hand, the ground state
$|\Psi_0^{\textrm{I}}(J_0,\gamma_0)\rangle\equiv
|\Psi(J_0,\gamma_0,V_s=0)\rangle_0$ of the noninteracting model with
hopping parameters $J_0$ and $\gamma_0$ as well as the ground state
$|\Psi_0^{\textrm{II}}(J_0,J'_0,\gamma_0)\rangle\equiv
|\Psi(J_0,J'_0,\gamma_0,V_s=0)\rangle_0$ of Eq.~(\ref{Model}) with an
additional hopping $J'_0$ between the bath sites. In order to study
the time evolution under the interacting Hamiltonian we use a
time-dependent density-matrix renormalization group (DMRG) algorithm
\cite{FeiguinWhite}, a method which has already been applied to study
other one-dimensional models
\cite{KollathLauchli,ManmanaWessel,HeidrichMeisnerRigol}. We choose
open boundary conditions with a chain length of $L=51$. The number of
states kept in the truncated adaptive Hilbert space varies between
$\chi=400$ and $\chi=800$. For a global quench as considered here it
is well known that the entanglement entropy between two subsystems usually
increases linearly with time. Since the maximal entanglement which can
be represented in a truncated Hilbert space is limited by $\ln\chi$,
there is a maximum time $t_{\rm max}$ up to which we can reliably
simulate the time evolution. For the cases considered here this time
scale is given by $Jt_{\rm max}\approx 15-25$.

\begin{figure}
\includegraphics*[width=0.8\columnwidth]{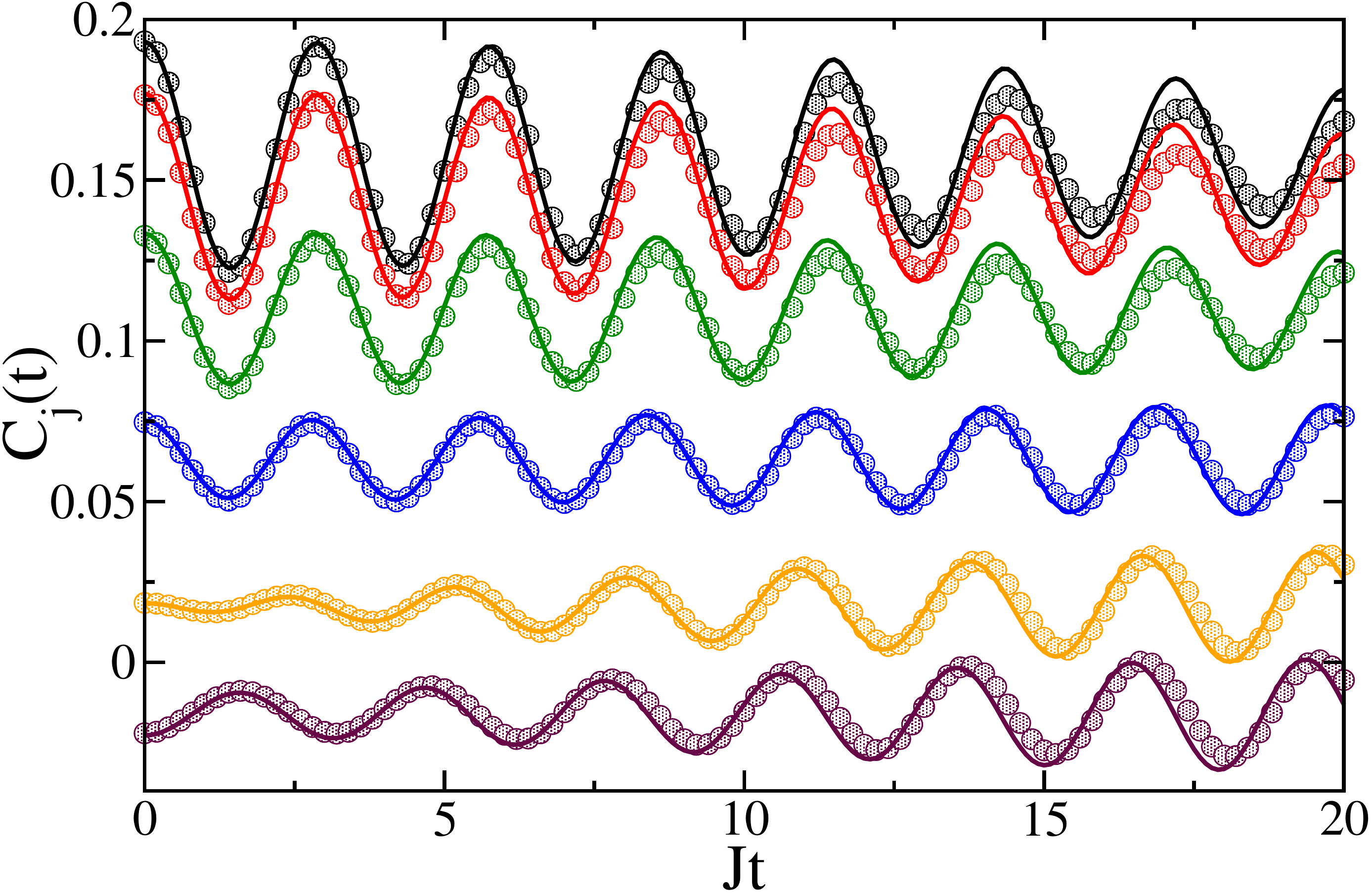}
\caption{(Color online) $C_j(t)$ from Eq.~\eqref{corr_fn} for
  $j=0,\ldots,5$ (top to bottom) for a quench with initial state
  $|\Psi_0^{\textrm{I}}(1,1)\rangle$, and Hamiltonian $H$ with
  $J=1$, $\gamma=1$, and $V_s=1$ at low densities. DMRG results
  (symbols) are compared to the solution of Eq.~\eqref{eq_motion}
  (lines).}
\label{Fig1}
\end{figure}

\paragraph{Results.}
First, we will concentrate on the relaxation dynamics at low particle
densities. As an example, we show in Fig.~\ref{Fig1} results for a
quantum quench with $N=11$ particles.
Shown are results for the one--point correlation functions
\begin{eqnarray}\label{corr_fn}
C_j(t)\equiv\langle\hat{C}_j\rangle_t=\langle\Psi(t)| c_{(L+1)/2}^\dagger c_{(L+1)/2+j}|\Psi(t)\rangle\;.
\end{eqnarray}
In all correlation functions oscillations with a characteristic
frequency are visible. These oscillations can be understood from an
equation of motion approach. We define the three time-dependent
expectation values $f_q(t)=\langle\Psi(t)|c^\dagger_q
c_q|\Psi(t)\rangle$, $g_q(t)=\langle\Psi(t)|s^\dagger_q
s_q|\Psi(t)\rangle$, and $\rho_q(t)=\langle\Psi(t)|c^\dagger_q
s_q|\Psi(t)\rangle$ where $c_q=\sqrt{2/(L+1)}\sum_j \sin(qj) c_j$ with
allowed momenta $q=n\pi/(L+1)$, and $n=1,\ldots,L$. Then, using
Heisenberg's equation of motion, a Hartree-Fock decoupling of the
quartic terms, and the additional assumption of an instantaneous
dephasing \cite{SupplMat}, we find the following system of coupled
equations
\begin{eqnarray}
\label{eq_motion}
\dot{f}_q(t)&=&-\dot{g}_q(t)=2\gamma R_q(t)\; ,  \quad
\dot{r}_q(t)=B_q(t) R_q(t)\;,\nonumber \\
\dot{R}_q(t)&=&-\gamma[f_q(t)-g_q(t)]-B_q(t) r_q(t)\;,
\end{eqnarray}
with $\varepsilon_q=-2\cos q$, $r_q(t)=\Re\rho_q(t)$,
$R_q(t)=\Im\rho_q(t)$, and
$B_q(t)=-V_s-\varepsilon_q+2V_s/(L+1)\bigl[\cos^2(q)
g_{\pi-q}(t)-\sin^2(q) g_q(t) +\sum_k\bigl(1-\cos k\cos q\bigr)g_k(t)\bigr]\approx
-V_s-\varepsilon_q\equiv\mathcal{B}_q$.

We solve the set of Eqs.~(\ref{eq_motion}) numerically, and the
results up to intermediate times are in excellent agreement with the
DMRG data, see Fig.~\ref{Fig1}. Using further approximations, we
analytically find that the oscillation frequency is given by
$\Omega^2_q=\mathcal{B}_q^2+(2\gamma)^2$ with $\Omega^2_{q\to 0}\approx
1+(2\gamma)^2$ and depends only weakly on $q$
\cite{SupplMat}. This means that the dephasing process is very slow.
For longer times and short distances we see that the amplitude of the
oscillations in the DMRG data is decaying faster than predicted by our
equations of motion approach. Here it is important to realize that due
to the Hartree-Fock decoupling the equations of motion effectively
describe the time evolution under a free particle Hamiltonian. This
approach therefore takes only the slow dephasing process discussed
above into account.  The additional decay seen in the DMRG data is due
to slow relaxation processes involving energy-momentum transfer
between interacting particles which are not captured in our equations
of motion approach.

\begin{figure}
\includegraphics*[width=1.0\columnwidth]{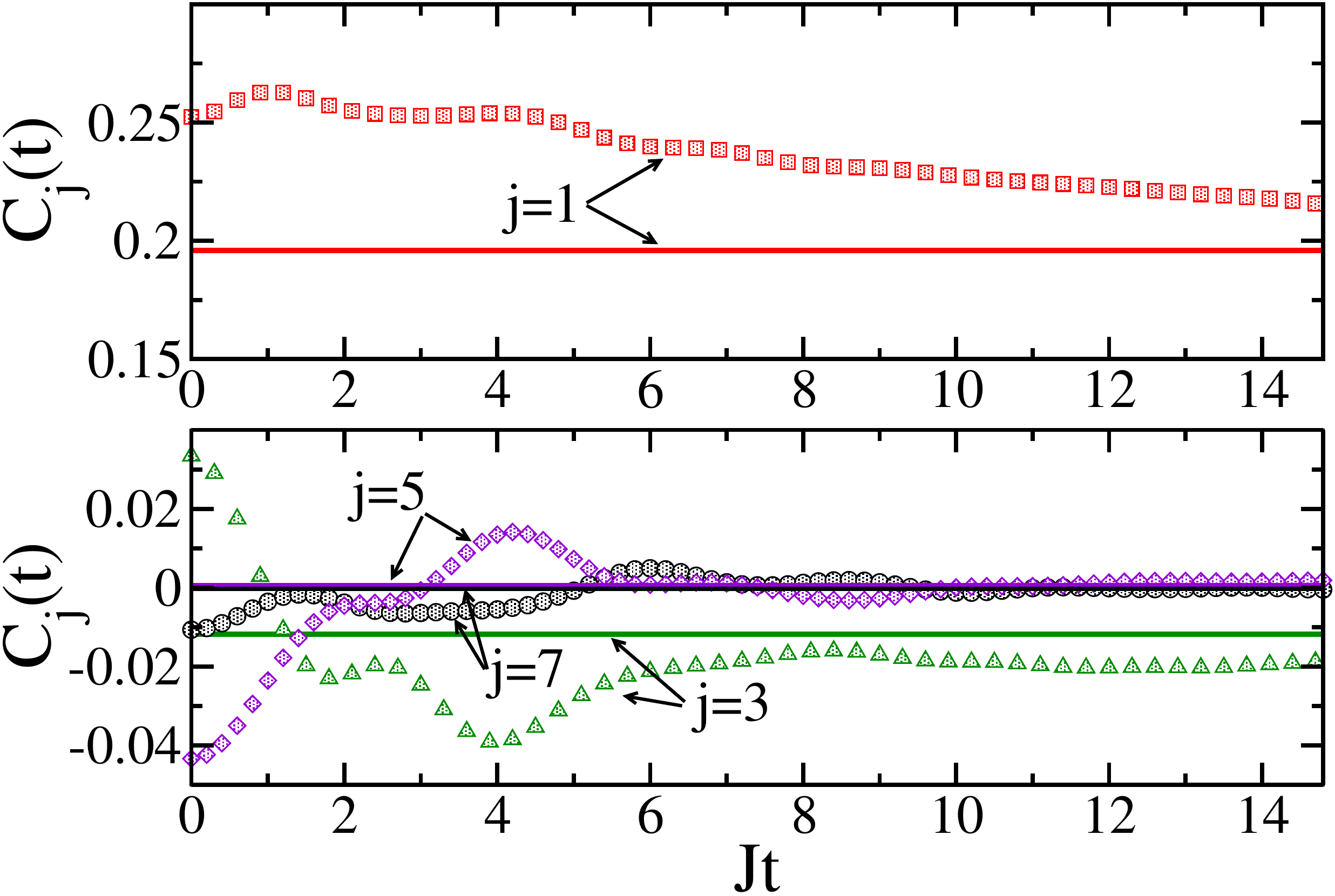}
\caption{(Color online) DMRG data (symbols) for $C_j(t)$ at
  half filing for a quench with initial state
  $|\Psi_0^{\textrm{II}}(1,0.6,1)\rangle$, and Hamiltonian $H$ with $J=1$,
  $\gamma=1$, and $V_s=1$.  The lines are the thermal expectation
  values $\langle C_j\rangle_T$.}
\label{Fig3}
\end{figure}

A much faster relaxation occurs if we increase the particle density
with a maximum in the relaxation rate at half filing. The DMRG data
for a quench in the half-filled case in Fig.~\ref{Fig3} show indeed
that the system almost completely equilibrates within the simulation
time $t_{\textrm{max}}$.
Due to the particle--hole symmetry of the Hamiltonian and the initial
state we have $C_0(t)\equiv 1/2$ and $C_{2j}(t)\equiv 0$. For odd
distances we now see, instead of long-time oscillations, an
exponential damping which allows us to extrapolate the correlation
functions and to read off the value for $C_j(t\to\infty)$
\cite{SupplMat}. Due to the lightcone-like spreading of the correlations
\cite{LiebRobinson,EnssSirker}, the short-range correlation functions
in the middle of the chain are, for the time range shown in
Fig.~\ref{Fig3}, not affected by the boundaries and are almost
indistinguishable from those for an infinite system. By extrapolating
our numerical data we thus approximately obtain $C_j(t\to\infty)$ for
a system in the thermodynamic limit.

\begin{figure}
\includegraphics*[width=1.0\columnwidth]{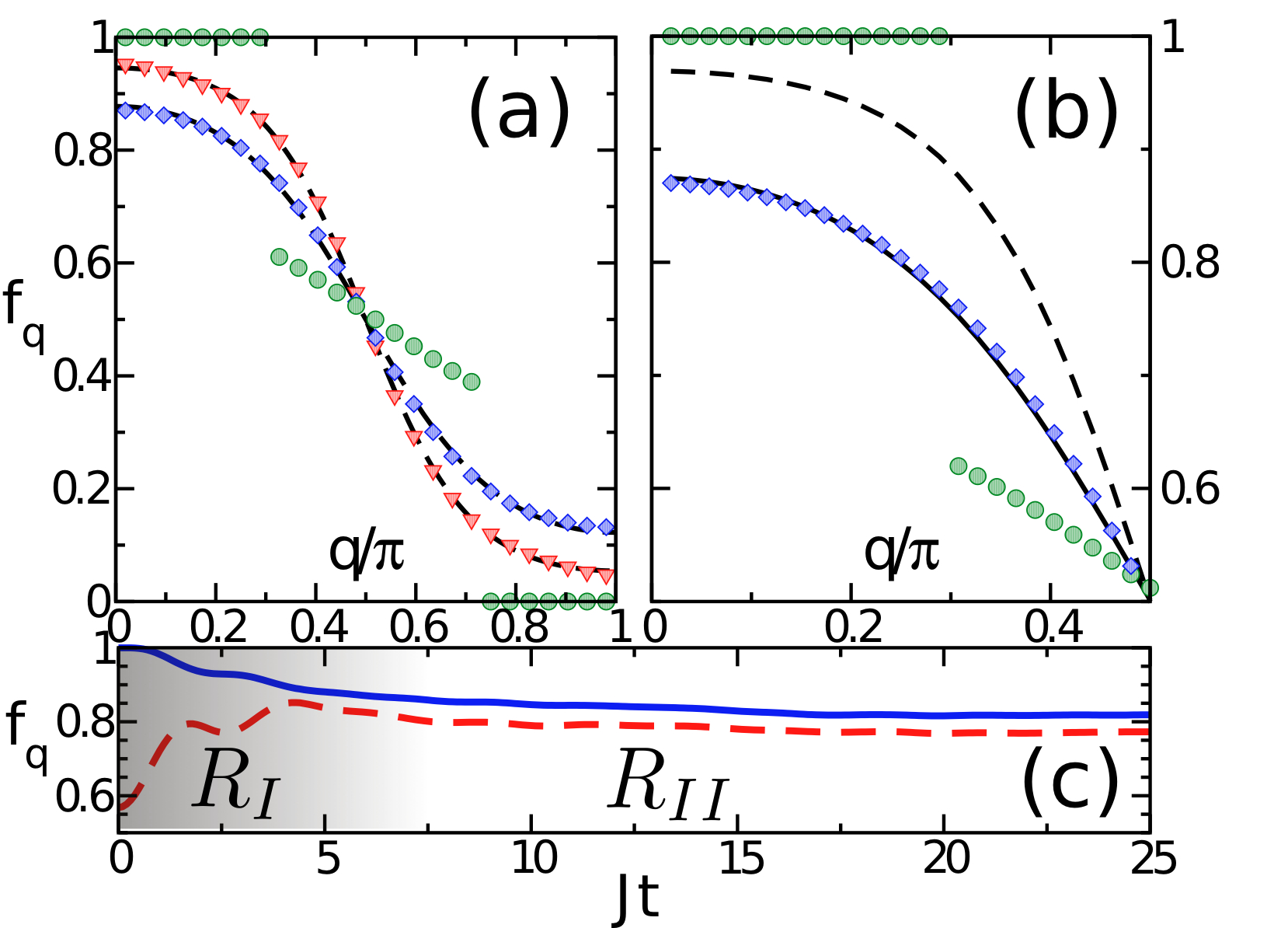}
\caption{(Color online) (a) $f_q(t=0)$ for
  $|\Psi_0^{\textrm{II}}(1,0.6,1)\rangle$ (circles), $f_q(t=5)$
  (triangles) and Fermi function fit $\langle f_q\rangle_{T=0.7J}$,
  and the extrapolated distribution $f_q(t\to\infty)$ (diamonds) with
  a fit $\langle f_q\rangle_{T=J}$. (b) $f_q(t\to\infty)$ (diamonds)
  compared to the thermal average $\Tr\{f_q\e^{-H/T}\}/Z$ (solid line)
  and $\langle f_q\rangle_{T}$ (dashed line) where $T/J= 0.54$ is
  fixed by Eq.~(\ref{Lagrange}).  (c) $f_q(t)$ for $q=13\pi/52$ (solid
  line) and $q=16\pi/52$ (dashed line).}
\label{Fig4}
\end{figure}

The corresponding distribution function $f_q(t)$, shown in
Fig.~\ref{Fig4}(a), has already become completely smooth after a short
time, $Jt=5$, and can be well fitted by a free fermion
distribution function $\langle
f_q\rangle_{T}=1/(e^{\varepsilon_q/T}+1)$.
However, the system has not fully equilibrated yet. Fig.~\ref{Fig4}(c)
shows that we have two distinct relaxation regimes. In regime $R_I$ we
have a relatively quick reshuffling in the distribution leading to a
prethermalized state \cite{RigolDunjko,GringKuhnert}. This is followed
by a slow drift of the occupation numbers in regime $R_{II}$ which,
when extrapolated in time, leads to the final distribution for the
equilibrated state. While both distributions can be well fitted by
$\langle f_q\rangle_T$, the temperature should not be used as a
fitting parameter but should rather be determined by energy conservation. We
therefore expect that the equilibrated system is described by the
ensemble, $\rho=\exp(-H/T)/Z$, with the chemical potential $\mu=0$ due to
particle--hole symmetry. The temperature T is determined by
Eq.~(\ref{Lagrange}) with $\mathcal{Q}_n$ replaced by $H$. The lhs of
Eq.~\eqref{Lagrange} is now an expectation value for a noninteracting
system and can be obtained analytically. The thermal average on the
rhs is calculated using a static DMRG calculation \cite{FeiguinWhite}.
For the particular quench in
Fig.~\ref{Fig4} we find $T/J= 0.54$. This then allows the calculation
of $\langle C_j\rangle_T\equiv \Tr\{\hat{C}_j\e^{-H/T}\}/Z$ by the DMRG algorithm
as shown in Fig.~\ref{Fig3}. The results for the corresponding
distribution function are shown as a solid line in Fig.~\ref{Fig4}(b)
and agree well with the time extrapolated values, demonstrating a
local thermalization.

\begin{figure}
\includegraphics*[width=1.0\columnwidth]{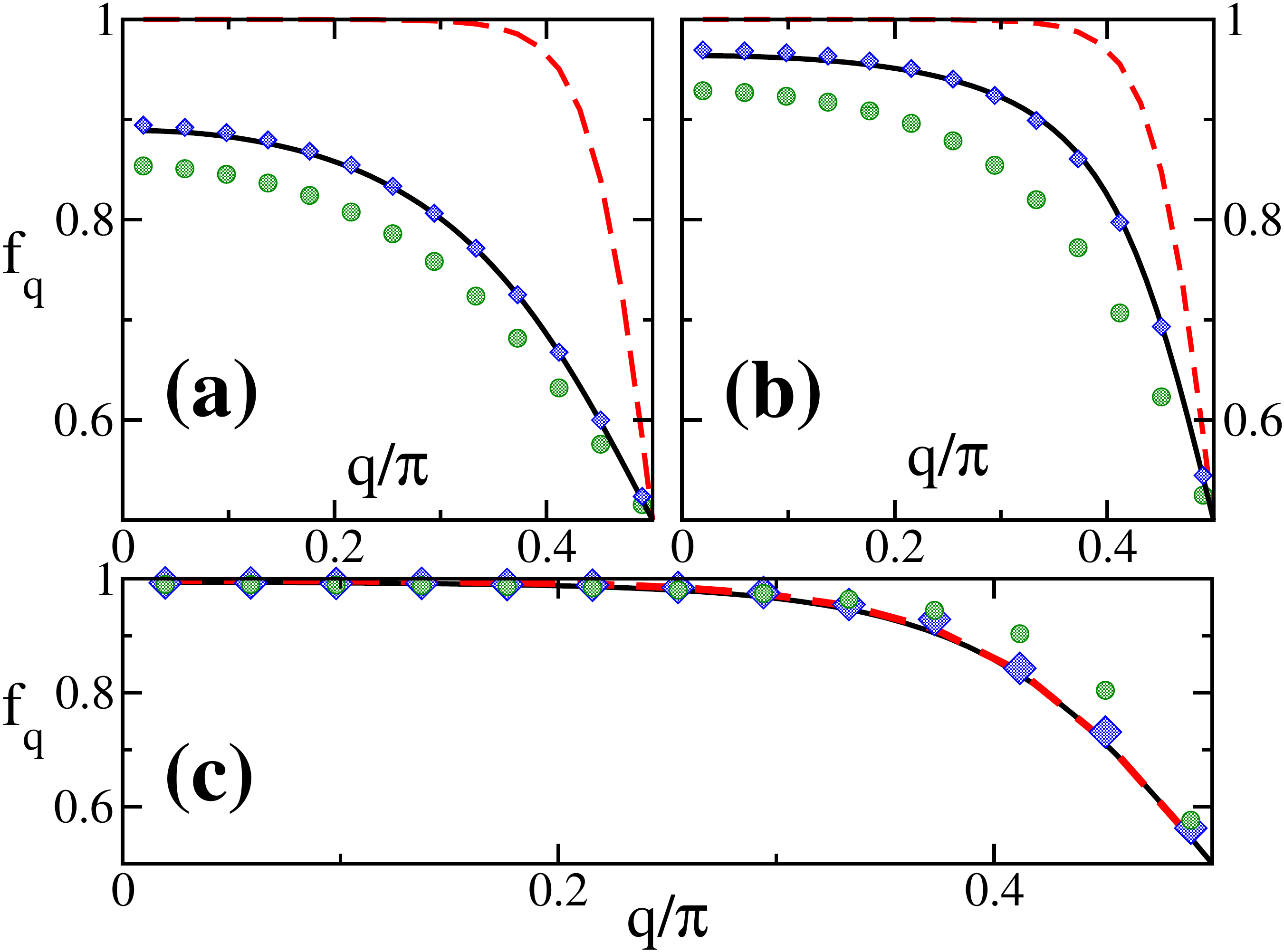}
\caption{(Color online) The initial distribution (circles),
  $f_q(t\to\infty)$ from DMRG (diamonds), the thermal distribution
  $\Tr\{f_q\e^{-H/T}\}/Z$ (solid line), and the free fermion
  distribution $\langle f_q\rangle_T$ (dashed line) with: (a)
  $\gamma=1$, $T/J=0.19$, (b) $\gamma=0.6$, $T/J=0.18$, and (c)
  $\gamma=0.2$, $T/J=0.33$.}
\label{Fig5}
\end{figure}

If the additional sites are to represent an effective bath, the
distribution function in the chain should become a Fermi distribution.
However, as can be seen in Fig.~\ref{Fig4}(b), $\langle
f_q\rangle_{T=0.54J}$ differs significantly from the equilibrium
distribution. One obvious reason is that the effective coupling
between the chain and bath in the thermal state $\sim \gamma \langle
s^\dagger_i c_i\rangle_{T=0.54J}\approx 0.28\gamma$ is not small.
Next, we therefore consider cases where we successively reduce the
coupling $\gamma$. In order to be able to still find the equilibrated
state within the limited simulation time we now use as the initial state
$|\Psi_0^{\textrm{I}}(1,\gamma)\rangle$ which yields a much smoother initial
distribution. Results for different coupling strengths $\gamma$ are
shown in Fig.~\ref{Fig5}.
We indeed find that the momentum distribution in equilibrium now
approaches the free fermion distribution with the temperature
determined by Eq.~(\ref{Lagrange}). At $\gamma=0.2$ the effective
coupling between the chain and bath $\propto \gamma \langle s^\dagger_i
c_i\rangle_{T=0.33J}\approx 0.06\gamma$ is very small.
Apart from the usual Pauli blocking there is another mechanism which
explains the very weak coupling between the subsystems.  Because the
nearest neighbor occupation $\langle
n_j^Bn_{j+1}^B\rangle_{T=0.33J}=0.1$ is also small we can
approximately project out all states where nearest-neighbor sites in
the bath are occupied.
This leads to an effective density-density interaction $\propto
(\gamma^2/V_s) n_j^A n_j^B$ between the subsystems, leading to a slow
relaxation for small $\gamma$ \cite{SupplMat}. In this strong coupling
limit, the hybridization part of the Hamiltonian Eq.~(\ref{Model}) also
gets projected $\propto \gamma
(s^\dagger_jc_j+h.c.)(1-n_{j-1}^B)(1-n_{j+1}^B)$ explaining the small
value for the effective coupling given above. Thus the interactions
help to decouple the two subsystems explaining the almost perfect free
Fermi distribution in the chain for $\gamma=0.2$.

\paragraph{Conclusions.}
We have studied thermalization in a strongly correlated model which
can be viewed either as a closed quantum system or as a free fermionic
chain coupled to a bath. Contrary to the common approach of using a
Lindblad equation to study open quantum systems, our model has a
microscopically specified bath. Therefore we can simulate the
nonequilibrium dynamics of the system {\it and} bath, and directly compare
the two different viewpoints. For low particle densities we have shown
that an equation of motion approach on the Hartree-Fock level is
sufficient to quantitatively describe the intermediate time dynamics.
At this level only slow dephasing processes are captured. For the
future it seems promising to use a higher order decoupling which might
also capture the faster relaxation processes which we observe in the
numerical simulations.  While the relaxation rate $\Gamma\sim
V_s\langle n_j^Bn_{j+1}^B\rangle$ at small interactions or low
densities is too small to observe equilibration within the limited
numerical simulation time we do observe thermalization at stronger
interactions near half filing where $\Gamma$
is larger. We note that the relaxation rate changes continuously with
the microscopic parameters of the model so that the definition of a
`nonequilibrium phase transition' based on the accessible simulation
time $t_{\rm max}$ seems problematic \cite{KollathLauchli}. Most
interestingly, we find that strong interactions lead to an effective
disentanglement between the subsystems and therefore increase
the decoherence times. Furthermore, even an extremely simple bath where
Markovian dynamics cannot be taken for granted can be sufficient to
fully equilibrate a subsystem without intrinsic relaxation processes.

\acknowledgments The authors thank S.~Manmana, F.H.L.~Essler, and L.~Santos
for discussions. J.S.~and N.S.~acknowledge support by the
Collaborative Research Centre SFB/TR49 and the graduate
school of excellence MAINZ and J.R.~acknowledges support by the National Natural Science Foundation of
China (Grant No.~11104021).

\appendix
\newpage
\section{Equation of motion approach}

We set up a set of equations for the following three
time-dependent expectation values
\begin{eqnarray}
f_q(t)&=&\langle\Psi(t)|c^\dagger_qc_q|\Psi(t)\rangle\;,\nonumber\\
g_q(t)&=&\langle\Psi(t)|s^\dagger_qs_q|\Psi(t)\rangle\;\textrm{, and}\\\nonumber
\rho_q(t)&=&\langle\Psi(t)|c^\dagger_qs_q|\Psi(t)\rangle\;.
\end{eqnarray}
Using Heisenberg's equation of motion,
\begin{eqnarray}
 \dot{O}=\im[H,O]\;,
\end{eqnarray}
with the Hamiltonian given by Eq.~(1) of the main text and the standard fermionic commutation relations, one finds, with $n_j^B=s_j^\dagger s_j$,
\begin{widetext}\begin{eqnarray}
\dot{f}_q(t)&=&2\gamma R_q(t)\;,\\
\dot{g}_q(t)&=&-2\gamma R_q(t)+\frac{2V_s}{L+1}\sum_{i\neq j}\sin\left[qi\right]\sin\left[qj\right]
\langle\Psi(t)|\left(s^\dagger_is_j-s^\dagger_js_i\right)
\left[n^B_{j+1}\left(1-\delta_{i,j+1}\right)+n^B_{j-1}\left(1-\delta_{i,j-1}\right)\right]|\Psi(t)\rangle\;,\nonumber\\
\im\dot{\rho}_q(t)&=&\im[\dot r_q(t)+\im\dot R_q(t)]=-\left[\varepsilon_q+V_s\right]\rho_q(t)+\gamma\left[f_q(t)-g_q(t)\right]\nonumber\\&&
+V_s\sqrt{\frac{2}{L+1}}\left[\sum_{j=1}^{L-1}\sin\left[qj\right]\langle\Psi(t)|c^\dagger_qs_jn^B_{j+1}|\Psi(t)\rangle+
\sum_{j=2}^{L}\sin\left[qj\right]\langle\Psi(t)|c^\dagger_qs_jn^B_{j-1}|\Psi(t)\rangle\right]\;.\nonumber
\end{eqnarray}\end{widetext}
We have introduced the real functions $r_q(t)=\Re\rho_q(t)$ and
$R_q(t)=\Im\rho_q(t)$. This set of coupled equations is exact.

To solve this set of equations we apply two approximations. Firstly a
Hartree--Fock decoupling is used, {\it i.e.}~we apply Wick's theorem
so that we have only two point correlation functions present, for
example
\begin{eqnarray}
\langle c^\dagger_qs_jn^B_{j-1} \rangle&=&\langle c^\dagger_qs_j \rangle\langle n^B_{j-1} \rangle
-\langle c^\dagger_qs_{j-1} \rangle\langle s^\dagger_{j-1}s_{j} \rangle\;.
\end{eqnarray}
One should note that, amongst other effects, this approximation leaves
\begin{eqnarray}
\dot{g}_q(t)&=&-2\gamma R_q(t)=-\dot{f}_q(t)
\end{eqnarray}
and hence $g_q(t)+f_q(t)\equiv N_q$ becomes independent of time.
Therefore, by performing the Hartree--Fock decoupling, we lose all
relaxation processes which can reshuffle the occupation of the
momenta. Nonetheless, as demonstrated in Fig.~1 of the main text, for
small densities this approximation is sufficient to get good
quantitative agreement with DMRG calculations for intermediate times.
The second approximation is ``instantaneous dephasing'', which means
that all off-diagonal elements of
$\langle\Psi(t)|c^\dagger_qc_k|\Psi(t)\rangle$, {\it etc.}, are taken
to `instantaneously dephase' and we keep only diagonal terms. For a
periodic system this would be guaranteed by translational invariance,
here it amounts to disregarding finite size effects from the
boundaries. This means that all two point correlation functions are
diagonal in momentum space. Following this we have
\begin{eqnarray}\label{eqset}
\dot{f}_q(t)&=&-\dot{g}_q(t)=2\gamma R_q(t)\;,\nonumber\quad
\dot{r}_q(t)=B_q(t) R_q(t)\;,\\
\dot{R}_q(t)&=&-\gamma[f_q(t)-g_q(t)]-B_q(t)r_q(t)\;,
\end{eqnarray}
with
\begin{eqnarray}
\label{B_q}
 B_q(t)&=&\frac{2V_s}{L+1}\bigg[\cos^2 \left[q\right]g_{\pi-q}(t)-\sin^2 \left[q\right]g_q(t)\\&&+\sum_k\left(1-\cos\left[k\right]\cos\left[q\right]\right)g_k(t)\bigg]-V_s-\varepsilon_q\;.\nonumber
\end{eqnarray}
\begin{figure}
\includegraphics*[width=1.0\columnwidth]{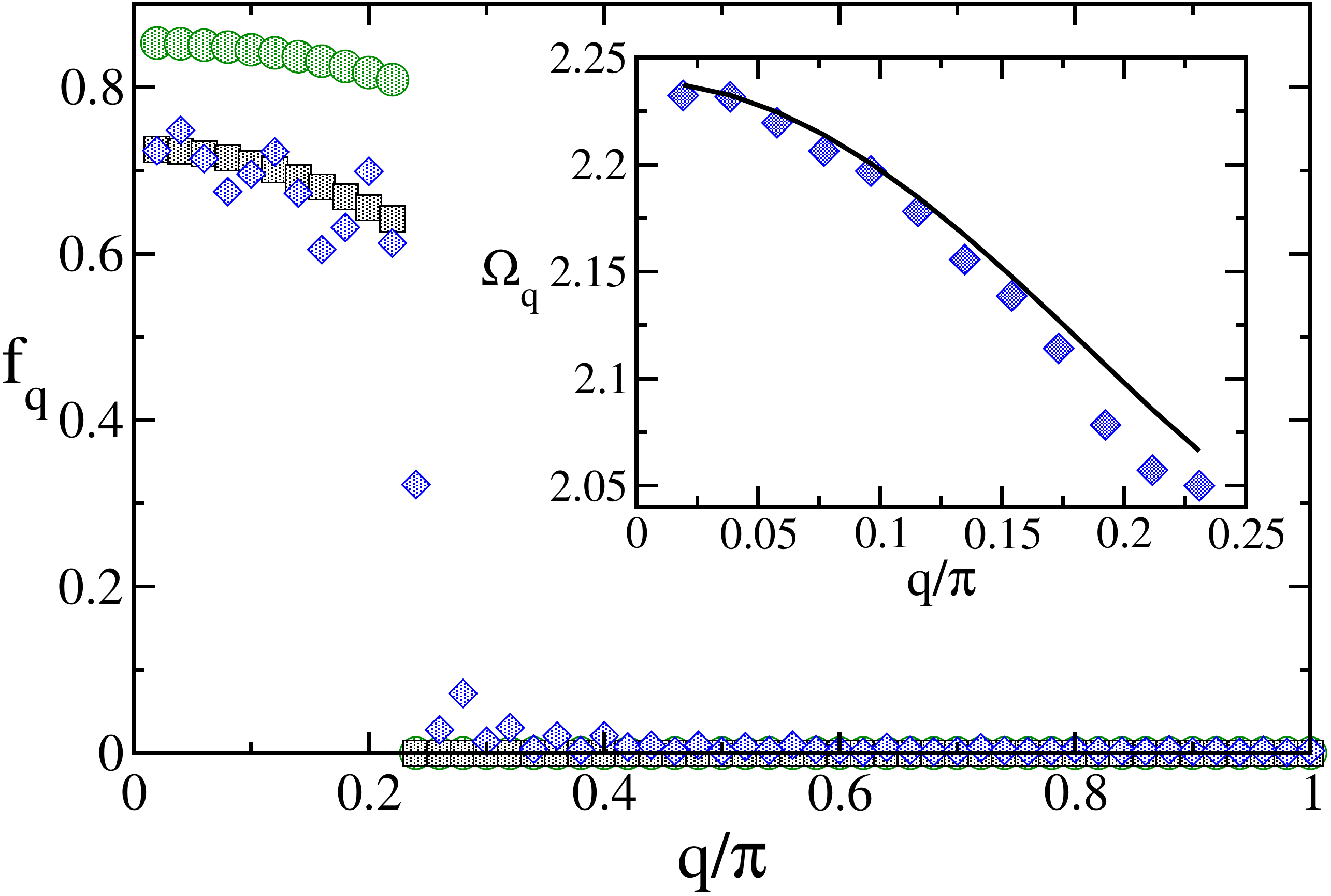}
\caption{Main: A quench with initial state
  $|\Psi_0^{\textrm{I}}(1,1)\rangle$, and Hamiltonian $H$ with $J=1$,
  $\gamma=1$, and $V_s=1$ at low densities. Shown is $f_q(t=0)$
  (circles), $\bar f_q$ within Hartree--Fock (squares), and $\bar f_q$
  obtained by DMRG (diamonds).  Inset: $\Omega_q$ obtained by DMRG
  (symbols) and within Hartree--Fock (line).}
\label{lowdensity_fig}
\end{figure}
The coupled first order differential equations, given by
Eq.~\eqref{eqset}, can then be solved iteratively. The first line in
Eq.~(\ref{B_q}) is a $\mathcal{O}(1/L)$ finite size correction.

The oscillation frequency of $R_q(t)$ is the same as that of $f_q(t)$
and $g_q(t)$ and can be extracted from these equations analytically.
We can write a second order differential equation for $R_q(t)$:
\begin{eqnarray}
\ddot{R}_q(t)+\left(4\gamma^2+B_q^2(t)\right)R_q(t)&=&-\dot{B}_q(t)r_q(t)\;.
\end{eqnarray}
One finds, with a weakly time dependent bath occupation, such that $\dot{B}_q\approx 0$,
\begin{eqnarray}
\ddot{R}_q(t)+\underbrace{\left(4\gamma^2+B_q^2(t)\right)}_{\equiv\Omega_q^2}R_q(t)&=&0\;.
\end{eqnarray}
 For small particle densities in the bath we can approximate $B_q(t)\approx -V_s-\varepsilon_q\equiv\mathcal{B}_q$ which gives
\begin{eqnarray}
\Omega_q\approx\sqrt{(\varepsilon_q+V_s)^2+4\gamma^2}\;.
\end{eqnarray}
The Hartree--Fock decoupled solutions oscillate with the frequency
$\Omega_q$ which is only weakly $q$--dependent, see inset of
Fig.~\ref{lowdensity_fig}. This explains why no dephasing effects are
seen in the Hartree--Fock solution on the timescales that we consider,
see Fig.~1 in the main text.  The approximation
$B_q(t)\approx\mathcal{B}_q$ is ensured in our case by the small bath
occupation. For example, in the initial state
$|\Psi_0^{\textrm{I}}(1,1)\rangle$, at a density of $0.11$ particles
per site, we have
\begin{eqnarray}
\frac{1}{L+1}\sum_q g_q(t=0)=0.0344\ll |{\cal B}_q|
\end{eqnarray}
and
\begin{eqnarray}
\frac{1}{L+1}\sum_q g_q(t=0)\cos q=0.0314\ll |{\cal B}_q|\,.
\end{eqnarray}

Contrary to the Hartree--Fock results, the DMRG data show an
additional relaxation, see Fig.~1 of the main text. A signature of the
beginning of this relaxation can also be seen in the long time mean of
the distribution function, $\bar{f}_q$, see Fig.~\ref{lowdensity_fig}
which shows a redistribution of the occupation of quasi-momenta around
the Fermi momentum.  The low density relaxation rate $\Gamma\sim
V_s\sum_j\langle n^B_jn^B_{j+1}\rangle/L$, however, is small so that
we can not see full thermalization within the DMRG simulation time
$t_{\textrm{max}}$.

\section{Particle--hole symmetry}

We define $H'$ as the Hamiltonian given by Eq.~(1) in the main text, but with hopping in the bath included:
\begin{eqnarray}
\label{Model_2}
H'&=&H -J'\sum_{j=1}^{L-1} \left \{ s_j^\dagger s_{j+1} +h.c.\right\}\;.
\end{eqnarray}
$H'$, and therefore also all half--filled groundstates, has particle--hole symmetry.
The Hamiltonian $H'$ is invariant under the mapping
\begin{eqnarray}\label{mapping}
{\cal T}_{\textrm{ph}}:\left\{\begin{array}{ccc}
c_j&\to&(-1)^jc_j^\dagger\\
c_j^\dagger&\to&(-1)^jc_j\\
s_j&\to&(-1)^{j+1}s_j^\dagger\\
s_j^\dagger&\to&(-1)^{j+1}s_j
       \end{array}\right.
\end{eqnarray}
which exactly describes particle--hole inversion.
\begin{figure}
\includegraphics*[width=1.0\columnwidth]{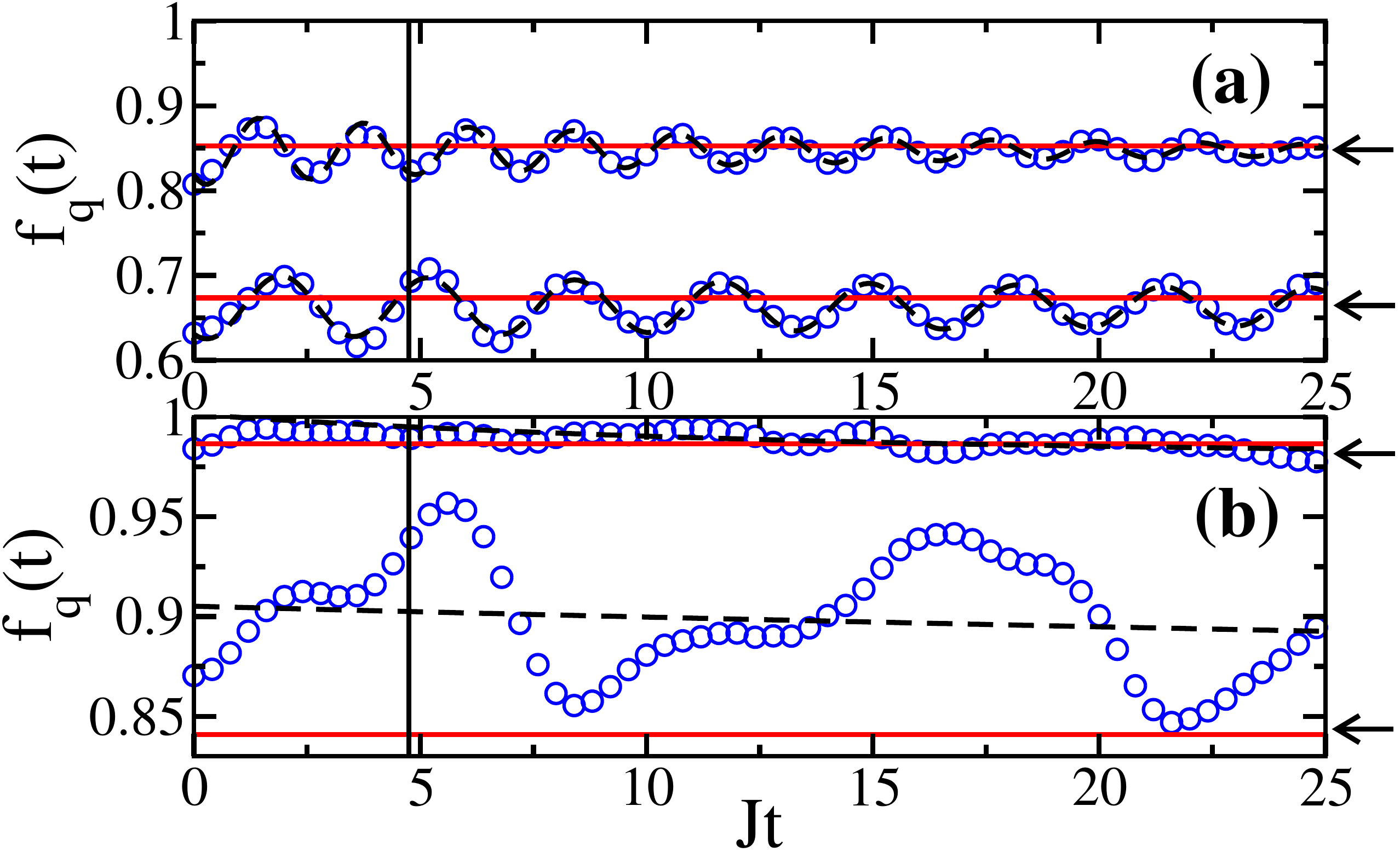}
\caption{Momentum space extrapolation for a quench with initial state
  $|\Psi_0^{\textrm{I}}(1,\gamma)\rangle$, and Hamiltonian $H$ with
  $J=1$ and $V_s=1$ where (a) $\gamma=1$, and (b) $\gamma=0.2$. Shown
  are the momenta $q=11\pi/(L+1)$ (upper curves) and $q=21\pi/(L+1)$
  (lower curves).  The dynamics (symbols) are compared with the fit
  (dashed line), and the thermal average at the appropriate
  temperature (solid line). Fitting is performed for times greater
  than the solid vertical line. The arrows on the right hand side show
  $f_q(t\to\infty)$ from (a) Eq.~(\ref{fit1}), and (b)
  Eq.~(\ref{fit2}).}
\label{Fit_I}
\end{figure}
\begin{figure}
\includegraphics*[width=1.0\columnwidth]{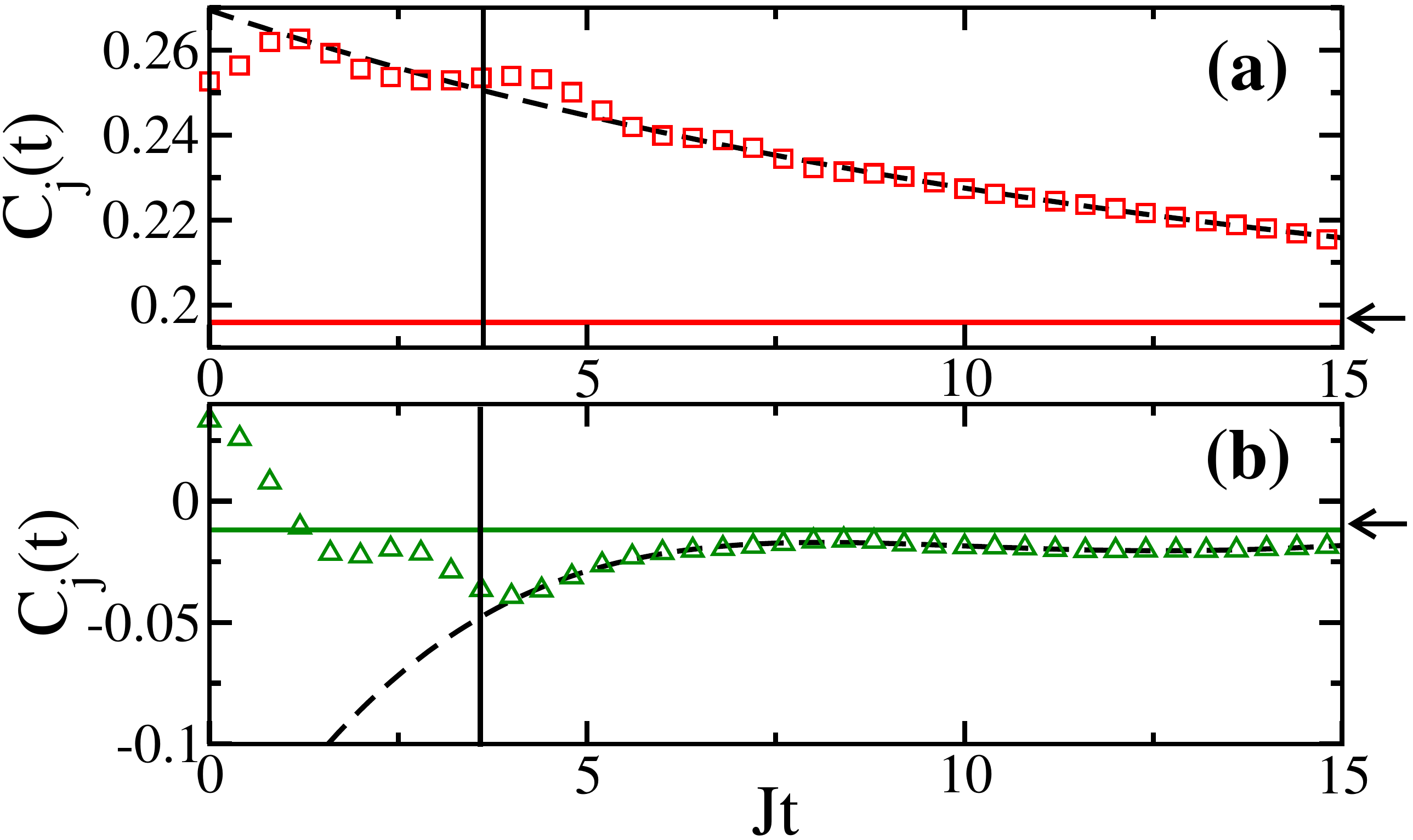}
\caption{Real space extrapolation for a quench with initial state
  $|\Psi_0^{\textrm{II}}(1,0.6,1)\rangle$, and Hamiltonian $H$ with
  $J=1$, $\gamma=1$, and $V_s=1$. The DMRG data (symbols) are compared
  with the fit (dashed line), and the thermal average at the
  appropriate temperature (solid line). Plotted is $C_j(t)$ with (a)
  $j=1$, and (b) $j=3$. Fitting is performed for times greater than
  the solid vertical line. The arrows on the right hand side show
  $C_j(t\to\infty)$ from Eq.~(\ref{fit3}).}
\label{Fit_II}
\end{figure}

In our analysis we have considered two correlation functions. Firstly the real space two point correlation function $C_j(t)$, defined by
\begin{eqnarray}
C_j(t)=\langle\Psi_0|e^{\im Ht} c_{(L+1)/2}^\dagger c_{(L+1)/2+j}e^{-\im Ht}|\Psi_0\rangle\;,
\end{eqnarray}
for time evolution with $H$ given by Eq.~(1) in the main text. Under the mapping given by Eq.~\eqref{mapping} $H\to H$ and $|\Psi_0\rangle\to|\Psi_0\rangle$ and one finds that
$C_{j=0}(t)=1/2$, and $C_{2j}(t)=0$ for non--zero $j$. Secondly we analyzed the momentum distribution in the chain, $f_q(t)$. For the initial states under consideration, and therefore for all times, this can be shown to satisfy $f_q(t)+f_{\pi-q}(t)=1$ by using the same mapping.

\section{Fitting and extrapolation}
\begin{figure}
  \includegraphics*[width=1.0\columnwidth]{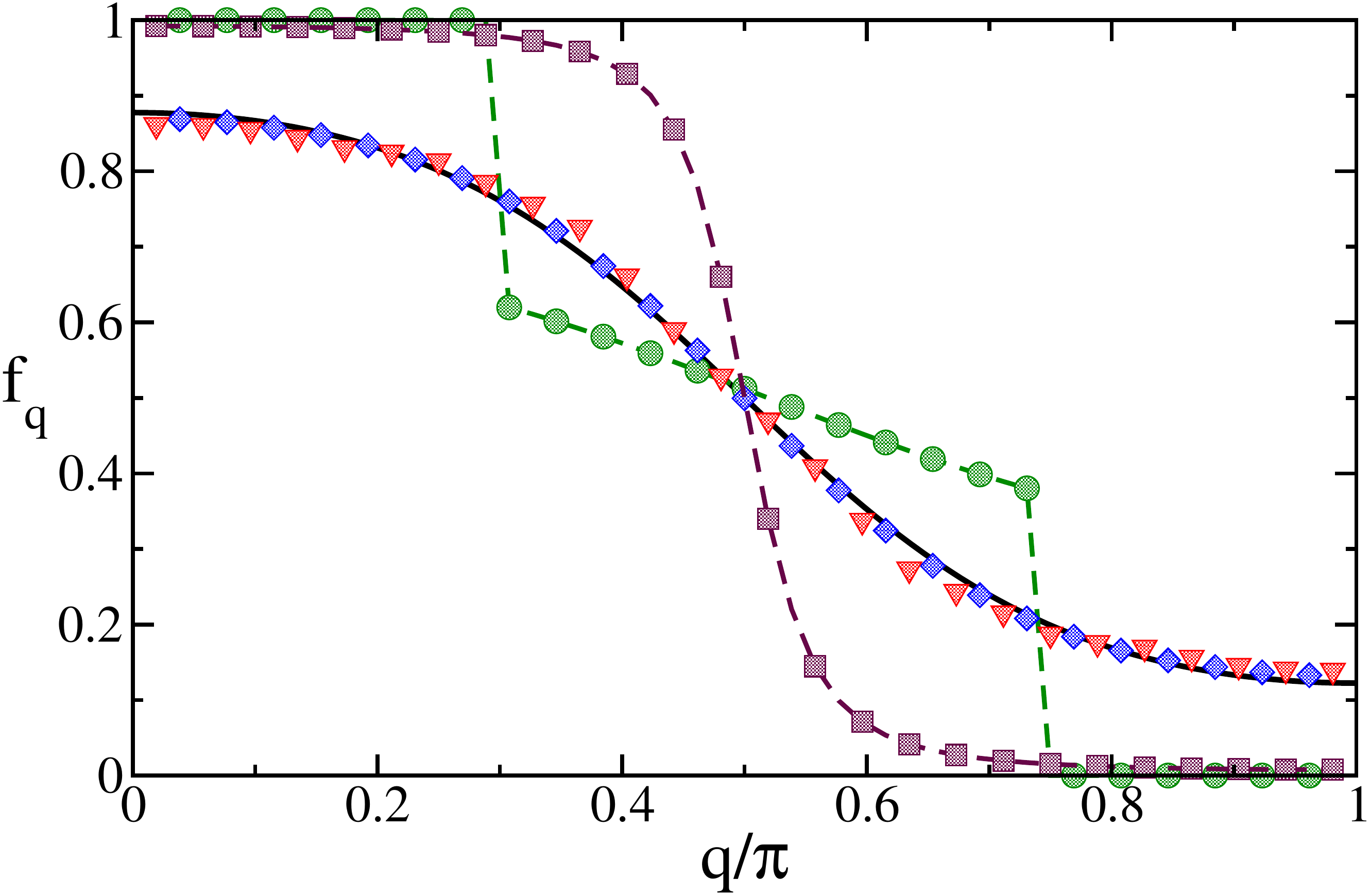}
  \caption{Distribution function $f_q$ in the chain for a system of
    size $L=51$. $f_q(t=0)$ is shown for the initial state
    $|\Psi_0^{\textrm{II}}(1,0.6,1)\rangle$ (circles), and the initial
    state $|\Psi_0^{\textrm{I}}(4.48,0.8)\rangle$ (squares). In both
    cases we time evolve with the same Hamiltonian $H$ with $J=1$,
    $\gamma=1$, and $V_s=1$. $f_q(t\to\infty)$ after time evolving
    $|\Psi_0^{II}\rangle$ (diamonds) and $|\Psi_0^{II}\rangle$
    (triangles) are compared with the thermal average
    $\Tr\{f_q\e^{-H/T}\}/Z$ (solid line) where $T/J=0.54$ is fixed by
    the initial energy, see text. Alternate $q$ points are plotted for
    the two quenches to aid clarity.}
\label{Fig_Quench_III}
\end{figure}
In this section we explain the fitting and extrapolation techniques we
used to find our long time data at half--filling. For quench I we
extrapolated in momentum space. The long time limit for $f_q$ and
$\gamma=1$ and $\gamma=0.6$ can be found directly by time averaging
the data or by fitting to
\begin{eqnarray}
\label{fit1}
  f_q(Jt\gg 1)\simeq f_q(t\to\infty)+a e^{-\Gamma t}\cos\left[\Omega t-\phi\right]\,.
\end{eqnarray}
This functional form takes into account exponential relaxation and a
simple oscillation, see Fig.~\ref{Fit_I}(a). Note that the fitting is
only performed right of the solid line at $Jt\approx 5$, to ignore the
effect of the short time dynamics. A Fourier analysis confirms that
there is one dominant frequency in the dynamics of $f_q(t)$. However,
in contrast to the low density case this frequency is momentum
dependent. In particular for the case $\gamma=1$ shown in
Fig.~\ref{Fit_I}(a), $\Omega(q=11\pi/(L+1))\approx 2.7$ and
$\Omega(q=21\pi/(L+1))\approx 1.9$. The relaxation rate,
$\Gamma(q=11\pi/(L+1))\approx 6.7\cdot 10^{-2}$ and
$\Gamma(q=21\pi/(L+1))\approx 2.7\cdot 10^{-2}$, is of the same order
of magnitude for all momenta hinting at one dominant relaxation
process. Let us stress that in this case the result for
$f_q(t\to\infty)$ depends only weakly on the extrapolation procedure
used, i.e. time averaging or fitting with different fit intervals,
with a variation in $f_q(t\to\infty)$ which is about the symbol size
used in the corresponding plots in the main text.

For $\gamma=0.2$, see Fig.~\ref{Fit_I}(b), such a simple fitting
function will no longer work due to the presence of various
oscillation frequencies. A Fourier analysis confirms that there is
more than one oscillation frequency involved, but the times are not
sufficient to extract how many there are and what their magnitudes may
be. Instead we trace out the overall trend by fitting to
\begin{eqnarray}
\label{fit2}
f_q(Jt\gg 1)\simeq f_q(t\to\infty)+a e^{-\tilde{\Gamma} t}\,.
\end{eqnarray}
$\tilde{\Gamma}$ captures the gradual drift of the oscillations which
can also be seen by using running averages. This procedure is robust
when choosing a variety of different time regions over which to
perform the fitting, and gives again errors smaller than the symbol
sizes used in the plots of the main text.

For quench II we extrapolate in real space. Fig.~\ref{Fit_II} shows
the fitting for $C_1(t)$ and $C_3(t)$. We fit the dynamics with
\begin{eqnarray}
\label{fit3}
C_j(Jt\gg 1)\simeq C_j(t\to\infty)+a e^{-\Gamma t}\left(\cos\left[\Omega t-\phi\right]+b\right).
\end{eqnarray}
As for quench I, only times right of the vertical lines in Fig.~\ref{Fit_II} are used for fitting.

\section{Initial state independence}
After relaxation the equilibrium state should depend only on the
energy in the system.  As example, we take two initial states,
$|\Psi_0^{\textrm{I}}\rangle$ and $|\Psi_0^{\textrm{II}}\rangle$,
constructed to have the same energy after a quench
\begin{eqnarray}
E&=&\langle\Psi^{\textrm{II}}_0|H|\Psi^{\textrm{II}}_0\rangle = \langle\Psi^{\textrm{I}}_0|H|\Psi^{\textrm{I}}_0\rangle\,.
\end{eqnarray}
The two different initial states are time evolved with the same
Hamiltonian, with $J=1$, $\gamma=1$, and $V_s=1$. We use the initial
states $|\Psi_0^{\textrm{II}}(1,0.6,1)\rangle$ (same as in Fig.~3 of
the main text) and $|\Psi_0^{\textrm{I}}(4.48,0.8)\rangle$ with energy
$E=-51.19$.  Fig.~\ref{Fig_Quench_III} demonstrates that both states
evolve towards the same equilibrium state, well described by the grand
canonical ensemble $\Tr\{f_q\e^{-H/T}\}/Z$ with the temperature
$T/J=0.54$ fixed by $E=\Tr\{H\e^{-H/T}\}/Z$ and $\mu=0$ due to
particle-hole symmetry.


\begin{thebibliography}{34}
\expandafter\ifx\csname natexlab\endcsname\relax\def\natexlab#1{#1}\fi
\expandafter\ifx\csname bibnamefont\endcsname\relax
  \def\bibnamefont#1{#1}\fi
\expandafter\ifx\csname bibfnamefont\endcsname\relax
  \def\bibfnamefont#1{#1}\fi
\expandafter\ifx\csname citenamefont\endcsname\relax
  \def\citenamefont#1{#1}\fi
\expandafter\ifx\csname url\endcsname\relax
  \def\url#1{\texttt{#1}}\fi
\expandafter\ifx\csname urlprefix\endcsname\relax\def\urlprefix{URL }\fi
\providecommand{\bibinfo}[2]{#2}
\providecommand{\eprint}[2][]{\url{#2}}

\bibitem[{\citenamefont{Breuer and Petruccione}(2002)}]{BreuerPetruccione}
\bibinfo{author}{\bibfnamefont{H.~P.} \bibnamefont{Breuer}} \bibnamefont{and}
  \bibinfo{author}{\bibfnamefont{F.}~\bibnamefont{Petruccione}},
  \emph{\bibinfo{title}{The Theory of Open Quantum Systems}}
  (\bibinfo{publisher}{Oxford University Press}, \bibinfo{address}{New York},
  \bibinfo{year}{2002}).

\bibitem[{\citenamefont{Weiss}(2008)}]{Weiss}
\bibinfo{author}{\bibfnamefont{U.}~\bibnamefont{Weiss}},
  \emph{\bibinfo{title}{Quantum Dissipative Systems}}
  (\bibinfo{publisher}{World Scientific}, \bibinfo{address}{Singapore},
  \bibinfo{year}{2008}), \bibinfo{edition}{3rd} ed.

\bibitem[{\citenamefont{Carmichael}(1998)}]{Carmichael}
\bibinfo{author}{\bibfnamefont{H.~P.} \bibnamefont{Carmichael}},
  \emph{\bibinfo{title}{Statistical Methods in Quantum Optics: Master Equations
  and Fokker--Planck Equations}} (\bibinfo{publisher}{Springer-Verlag},
\bibinfo{address}{Berlin}, \bibinfo{year}{1998}).

\bibitem[{\citenamefont{gebhardmuenster}(2011)}]{gebhardmuenster}
\bibinfo{author}{\bibfnamefont{F.} \bibnamefont{Gebhard}}
\bibnamefont{and}
  \bibinfo{author}{\bibfnamefont{K.} \bibnamefont{zu M\"unster}},
  \bibinfo{journal}{Ann.\ Phys.\ (Berlin)} \textbf{\bibinfo{volume}{523}},
  \bibinfo{pages}{552} (\bibinfo{year}{2011}).

\bibitem[{\citenamefont{Landau and Lifshitz}(1980)}]{LandauLifshitz3}
\bibinfo{author}{\bibfnamefont{L.~D.} \bibnamefont{Landau}} \bibnamefont{and}
  \bibinfo{author}{\bibfnamefont{E.~M.} \bibnamefont{Lifshitz}},
  \emph{\bibinfo{title}{Statistical Physics}} (\bibinfo{publisher}{Butterworth
  Heinemann}, \bibinfo{address}{Oxford},
\bibinfo{year}{1980}),
  \bibinfo{edition}{3rd} ed.

\bibitem[{\citenamefont{Rigol et~al.}(2008)\citenamefont{Rigol, Dunjko, and
  Olshanii}}]{RigolDunjko}
\bibinfo{author}{\bibfnamefont{M.}~\bibnamefont{Rigol}},
  \bibinfo{author}{\bibfnamefont{V.}~\bibnamefont{Dunjko}}, \bibnamefont{and}
  \bibinfo{author}{\bibfnamefont{M.}~\bibnamefont{Olshanii}},
  \bibinfo{journal}{Nature (London)} \textbf{\bibinfo{volume}{452}},
  \bibinfo{pages}{854} (\bibinfo{year}{2008}).

\bibitem[{\citenamefont{von Neumann}(1929)}]{vonNeumann}
\bibinfo{author}{\bibfnamefont{J.}~\bibnamefont{von Neumann}},
  \bibinfo{journal}{Z.\ Phys.} \textbf{\bibinfo{volume}{57}},
  \bibinfo{pages}{30} (\bibinfo{year}{1929}).

\bibitem[{\citenamefont{Trotzky et~al.}(2012)\citenamefont{Trotzky, Chen,
  Flesch, McCulloch, Schollw\"ock, Eisert, and Bloch}}]{TrotzkyChen}
\bibinfo{author}{\bibfnamefont{S.}~\bibnamefont{Trotzky}},
  \bibinfo{author}{\bibfnamefont{Y.-A.} \bibnamefont{Chen}},
  \bibinfo{author}{\bibfnamefont{A.}~\bibnamefont{Flesch}},
  \bibinfo{author}{\bibfnamefont{I.~P.} \bibnamefont{McCulloch}},
  \bibinfo{author}{\bibfnamefont{U.}~\bibnamefont{Schollw\"ock}},
  \bibinfo{author}{\bibfnamefont{J.}~\bibnamefont{Eisert}}, \bibnamefont{and}
  \bibinfo{author}{\bibfnamefont{I.}~\bibnamefont{Bloch}},
  \bibinfo{journal}{Nature\ Phys.} \textbf{\bibinfo{volume}{8}},
  \bibinfo{pages}{325} (\bibinfo{year}{2012}).

\bibitem[{\citenamefont{Kinoshita et~al.}(2006)\citenamefont{Kinoshita, Wenger,
  and Weiss}}]{KinoshitaWenger}
\bibinfo{author}{\bibfnamefont{T.}~\bibnamefont{Kinoshita}},
  \bibinfo{author}{\bibfnamefont{T.}~\bibnamefont{Wenger}}, \bibnamefont{and}
  \bibinfo{author}{\bibfnamefont{D.~S.} \bibnamefont{Weiss}},
  \bibinfo{journal}{Nature (London)} \textbf{\bibinfo{volume}{440}},
  \bibinfo{pages}{900} (\bibinfo{year}{2006}).

\bibitem[{\citenamefont{Hofferberth et~al.}(2007)\citenamefont{Hofferberth,
  Lesanovsky, Fischer, Schumm, and Schmiedmayer}}]{HofferberthLesanovsky}
\bibinfo{author}{\bibfnamefont{S.}~\bibnamefont{Hofferberth}},
  \bibinfo{author}{\bibfnamefont{I.}~\bibnamefont{Lesanovsky}},
  \bibinfo{author}{\bibfnamefont{B.}~\bibnamefont{Fischer}},
  \bibinfo{author}{\bibfnamefont{T.}~\bibnamefont{Schumm}}, \bibnamefont{and}
  \bibinfo{author}{\bibfnamefont{J.}~\bibnamefont{Schmiedmayer}},
  \bibinfo{journal}{Nature (London)} \textbf{\bibinfo{volume}{449}},
  \bibinfo{pages}{324} (\bibinfo{year}{2007}).

\bibitem[{\citenamefont{Strohmaier et~al.}(2010)\citenamefont{Strohmaier,
  Greif, J\"ordens, Tarruell, Moritz, Esslinger, Sensarma, Pekker, Altman, and
  Demler}}]{StrohmaierGreif}
\bibinfo{author}{\bibfnamefont{N.} \bibnamefont{Strohmaier}},
  \bibinfo{author}{\bibfnamefont{D.} \bibnamefont{Greif}},
  \bibinfo{author}{\bibfnamefont{R.} \bibnamefont{J\"ordens}},
  \bibinfo{author}{\bibfnamefont{L.} \bibnamefont{Tarruell}},
  \bibinfo{author}{\bibfnamefont{H.} \bibnamefont{Moritz}},
  \bibinfo{author}{\bibfnamefont{T.} \bibnamefont{Esslinger}},
  \bibinfo{author}{\bibfnamefont{R.} \bibnamefont{Sensarma}},
  \bibinfo{author}{\bibfnamefont{D.} \bibnamefont{Pekker}},
  \bibinfo{author}{\bibfnamefont{E.} \bibnamefont{Altman}}, \bibnamefont{and}
  \bibinfo{author}{\bibfnamefont{E.} \bibnamefont{Demler}},
  \bibinfo{journal}{Phys.\ Rev.\ Lett.} \textbf{\bibinfo{volume}{104}},
  \bibinfo{pages}{080401} (\bibinfo{year}{2010}).

\bibitem[{\citenamefont{Daley et~al.}(2004)\citenamefont{Daley, Kollath,
  Schollw\"ock, and Vidal}}]{DaleyKollath}
\bibinfo{author}{\bibfnamefont{A.~J.} \bibnamefont{Daley}},
  \bibinfo{author}{\bibfnamefont{C.}~\bibnamefont{Kollath}},
  \bibinfo{author}{\bibfnamefont{U.}~\bibnamefont{Schollw\"ock}},
  \bibnamefont{and} \bibinfo{author}{\bibfnamefont{G.}~\bibnamefont{Vidal}},
  \bibinfo{journal}{J.\ Stat.\ Mech.: Theor.\ Exp.} \bibinfo{pages}{P04005}
  (\bibinfo{year}{2004}).

\bibitem[{\citenamefont{White and Feiguin}(2004)}]{WhiteFeiguin}
\bibinfo{author}{\bibfnamefont{S.~R.}~\bibnamefont{White}} \bibnamefont{and}
  \bibinfo{author}{\bibfnamefont{A.~E.} \bibnamefont{Feiguin}},
  \bibinfo{journal}{Phys.\ Rev.\ Lett.} \textbf{\bibinfo{volume}{93}},
  \bibinfo{pages}{076401} (\bibinfo{year}{2004}).

\bibitem[{\citenamefont{Sirker and Kl\"umper}(2005)}]{SirkerKluemperDTMRG}
\bibinfo{author}{\bibfnamefont{J.}~\bibnamefont{Sirker}} \bibnamefont{and}
  \bibinfo{author}{\bibfnamefont{A.}~\bibnamefont{Kl\"umper}},
  \bibinfo{journal}{Phys.\ Rev.\ B} \textbf{\bibinfo{volume}{71}},
  \bibinfo{pages}{241101(R)} (\bibinfo{year}{2005}).

\bibitem[{\citenamefont{Vidal}(2003)}]{VidalTEBD1}
\bibinfo{author}{\bibfnamefont{G.}~\bibnamefont{Vidal}},
  \bibinfo{journal}{Phys.\ Rev.\ Lett.} \textbf{\bibinfo{volume}{91}},
  \bibinfo{pages}{147902} (\bibinfo{year}{2003}).

\bibitem[{\citenamefont{Vidal}(2004)}]{VidalTEBD2}
\bibinfo{author}{\bibfnamefont{G.}~\bibnamefont{Vidal}},
  \bibinfo{journal}{Phys.\ Rev.\ Lett.} \textbf{\bibinfo{volume}{93}},
  \bibinfo{pages}{040502} (\bibinfo{year}{2004}).

\bibitem[{\citenamefont{Enss and Sirker}(2012)}]{EnssSirker}
\bibinfo{author}{\bibfnamefont{T.}~\bibnamefont{Enss}} \bibnamefont{and}
  \bibinfo{author}{\bibfnamefont{J.}~\bibnamefont{Sirker}},
  \bibinfo{journal}{New J.\ Phys.} \textbf{\bibinfo{volume}{14}},
  \bibinfo{pages}{023008} (\bibinfo{year}{2012}).

\bibitem[{\citenamefont{Kehrein}(2005)}]{Kehrein}
\bibinfo{author}{\bibfnamefont{S.}~\bibnamefont{Kehrein}},
  \bibinfo{journal}{Phys.\ Rev.\ Lett.} \textbf{\bibinfo{volume}{95}},
  \bibinfo{pages}{056602} (\bibinfo{year}{2005}).

\bibitem[{\citenamefont{Anders and Schiller}(2005)}]{AndersSchiller}
\bibinfo{author}{\bibfnamefont{F.~B.} \bibnamefont{Anders}} \bibnamefont{and}
  \bibinfo{author}{\bibfnamefont{A.}~\bibnamefont{Schiller}},
  \bibinfo{journal}{Phys.\ Rev.\ Lett.} \textbf{\bibinfo{volume}{95}},
  \bibinfo{pages}{196801} (\bibinfo{year}{2005}).

\bibitem[{\citenamefont{Rigol et~al.}(2006)\citenamefont{Rigol, Muramatsu, and
  Olshanii}}]{RigolMuramatsu}
\bibinfo{author}{\bibfnamefont{M.}~\bibnamefont{Rigol}},
  \bibinfo{author}{\bibfnamefont{A.}~\bibnamefont{Muramatsu}},
  \bibnamefont{and} \bibinfo{author}{\bibfnamefont{M.}~\bibnamefont{Olshanii}},
  \bibinfo{journal}{Phys.\ Rev.\ A} \textbf{\bibinfo{volume}{74}},
  \bibinfo{pages}{053616} (\bibinfo{year}{2006}).

\bibitem[{\citenamefont{Kollath et~al.}(2007)\citenamefont{Kollath, L\"auchli,
  and Altman}}]{KollathLauchli}
\bibinfo{author}{\bibfnamefont{C.}~\bibnamefont{Kollath}},
  \bibinfo{author}{\bibfnamefont{A.~M.} \bibnamefont{L\"auchli}},
  \bibnamefont{and} \bibinfo{author}{\bibfnamefont{E.}~\bibnamefont{Altman}},
  \bibinfo{journal}{Phys.\ Rev.\ Lett.} \textbf{\bibinfo{volume}{98}},
  \bibinfo{pages}{180601} (\bibinfo{year}{2007}).

\bibitem[{\citenamefont{Manmana et~al.}(2007)\citenamefont{Manmana, Wessel,
  Noack, and Muramatsu}}]{ManmanaWessel}
\bibinfo{author}{\bibfnamefont{S.~R.} \bibnamefont{Manmana}},
  \bibinfo{author}{\bibfnamefont{S.}~\bibnamefont{Wessel}},
  \bibinfo{author}{\bibfnamefont{R.~M.} \bibnamefont{Noack}}, \bibnamefont{and}
  \bibinfo{author}{\bibfnamefont{A.}~\bibnamefont{Muramatsu}},
  \bibinfo{journal}{Phys.\ Rev.\ Lett.} \textbf{\bibinfo{volume}{98}},
  \bibinfo{pages}{210405} (\bibinfo{year}{2007}).

\bibitem[{\citenamefont{Biroli et~al.}(2010)\citenamefont{Biroli, Kollath, and
  L\"auchli}}]{BiroliKollath}
\bibinfo{author}{\bibfnamefont{G.}~\bibnamefont{Biroli}},
  \bibinfo{author}{\bibfnamefont{C.}~\bibnamefont{Kollath}}, \bibnamefont{and}
  \bibinfo{author}{\bibfnamefont{A.~M.} \bibnamefont{L\"auchli}},
  \bibinfo{journal}{Phys.\ Rev.\ Lett.} \textbf{\bibinfo{volume}{105}},
  \bibinfo{pages}{250401} (\bibinfo{year}{2010}).

\bibitem[{\citenamefont{Essler et~al.}(2005)\citenamefont{Essler, Frahm,
  G\"ohmann, Kl\"umper, and Korepin}}]{HubbardBook}
\bibinfo{author}{\bibfnamefont{F.~H.~L.} \bibnamefont{Essler}},
  \bibinfo{author}{\bibfnamefont{H.}~\bibnamefont{Frahm}},
  \bibinfo{author}{\bibfnamefont{F.}~\bibnamefont{G\"ohmann}},
  \bibinfo{author}{\bibfnamefont{A.}~\bibnamefont{Kl\"umper}},
  \bibnamefont{and} \bibinfo{author}{\bibfnamefont{V.~E.}
  \bibnamefont{Korepin}}, \emph{\bibinfo{title}{The One-Dimensional Hubbard
  Model}} (\bibinfo{publisher}{Cambridge University Press, Cambridge},
  \bibinfo{year}{2005}).

\bibitem[{\citenamefont{Sirker}(2012)}]{SirkerLL}
\bibinfo{author}{\bibfnamefont{J.}~\bibnamefont{Sirker}},
  \bibinfo{journal}{Int.\ J.\ Mod.\ Phys.\ B} \textbf{\bibinfo{volume}{26}},
  \bibinfo{pages}{1244009} (\bibinfo{year}{2012}).

\bibitem[{\citenamefont{Rigol et~al.}(2007)\citenamefont{Rigol, Dunjko,
  Yurovsky, and Olshanii}}]{RigolDunjkoPRL}
\bibinfo{author}{\bibfnamefont{M.}~\bibnamefont{Rigol}},
  \bibinfo{author}{\bibfnamefont{V.}~\bibnamefont{Dunjko}},
  \bibinfo{author}{\bibfnamefont{V.}~\bibnamefont{Yurovsky}}, \bibnamefont{and}
  \bibinfo{author}{\bibfnamefont{M.}~\bibnamefont{Olshanii}},
  \bibinfo{journal}{Phys.\ Rev.\ Lett.} \textbf{\bibinfo{volume}{98}},
  \bibinfo{pages}{050405} (\bibinfo{year}{2007}).

\bibitem[{\citenamefont{Rigol}(2009{\natexlab{a}})}]{Rigol}
\bibinfo{author}{\bibfnamefont{M.}~\bibnamefont{Rigol}},
  \bibinfo{journal}{Phys.\ Rev.\ Lett.} \textbf{\bibinfo{volume}{103}},
  \bibinfo{pages}{100403} (\bibinfo{year}{2009}{\natexlab{a}}).

\bibitem[{\citenamefont{Rigol}(2009{\natexlab{b}})}]{RigolPRA}
\bibinfo{author}{\bibfnamefont{M.}~\bibnamefont{Rigol}},
  \bibinfo{journal}{Phys.\ Rev.\ A} \textbf{\bibinfo{volume}{80}},
  \bibinfo{pages}{053607} (\bibinfo{year}{2009}{\natexlab{b}}).

\bibitem[{\citenamefont{Santos and Rigol}(2010)}]{SantosRigol}
\bibinfo{author}{\bibfnamefont{L.~F.} \bibnamefont{Santos}} \bibnamefont{and}
  \bibinfo{author}{\bibfnamefont{M.}~\bibnamefont{Rigol}},
  \bibinfo{journal}{Phys.\ Rev.\ E} \textbf{\bibinfo{volume}{81}},
  \bibinfo{pages}{036206} (\bibinfo{year}{2010}).

\bibitem[{\citenamefont{Gebhard et~al.}(2012)\citenamefont{Gebhard,
  zu~M\"unster, Ren, Sedlmayr, Sirker, and Ziebarth}}]{SirkerGebhard}
\bibinfo{author}{\bibfnamefont{F.} \bibnamefont{Gebhard}},
  \bibinfo{author}{\bibfnamefont{K.} \bibnamefont{zu~M\"unster}},
  \bibinfo{author}{\bibfnamefont{J.} \bibnamefont{Ren}},
  \bibinfo{author}{\bibfnamefont{N.} \bibnamefont{Sedlmayr}},
  \bibinfo{author}{\bibfnamefont{J.} \bibnamefont{Sirker}}, \bibnamefont{and}
  \bibinfo{author}{\bibfnamefont{B.} \bibnamefont{Ziebarth}},
  \bibinfo{journal}{Ann.\ Phys.\ (Berlin)} \textbf{\bibinfo{volume}{524}},
  \bibinfo{pages}{286} (\bibinfo{year}{2012}).

\bibitem[{\citenamefont{Feiguin and White}(2005)}]{FeiguinWhite}
\bibinfo{author}{\bibfnamefont{A.~E.} \bibnamefont{Feiguin}} \bibnamefont{and}
  \bibinfo{author}{\bibfnamefont{S.~R.} \bibnamefont{White}},
  \bibinfo{journal}{Phys.\ Rev.\ B} \textbf{\bibinfo{volume}{72}},
  \bibinfo{pages}{220401(R)} (\bibinfo{year}{2005}).

\bibitem[{\citenamefont{Heidrich-Meisner
  et~al.}(2008)\citenamefont{Heidrich-Meisner, Rigol, Muramatsu, Feiguin, and
  Dagotto}}]{HeidrichMeisnerRigol}
\bibinfo{author}{\bibfnamefont{F.} \bibnamefont{Heidrich-Meisner}},
  \bibinfo{author}{\bibfnamefont{M.} \bibnamefont{Rigol}},
  \bibinfo{author}{\bibfnamefont{A.} \bibnamefont{Muramatsu}},
  \bibinfo{author}{\bibfnamefont{A.~E.} \bibnamefont{Fei\-guin}},
  \bibnamefont{and} \bibinfo{author}{\bibfnamefont{E.} \bibnamefont{Dagotto}},
  \bibinfo{journal}{Phys.\ Rev.\ A} \textbf{\bibinfo{volume}{78}},
  \bibinfo{pages}{013620} (\bibinfo{year}{2008}).

\bibitem[{Sup()}]{SupplMat}
\bibinfo{note}{See appendix.}

\bibitem[{\citenamefont{Lieb and Robinson}(1972)}]{LiebRobinson}
\bibinfo{author}{\bibfnamefont{E.~H.} \bibnamefont{Lieb}} \bibnamefont{and}
  \bibinfo{author}{\bibfnamefont{D.~W.} \bibnamefont{Robinson}},
  \bibinfo{journal}{Commun.\ Math.\ Phys.} \textbf{\bibinfo{volume}{28}},
  \bibinfo{pages}{251} (\bibinfo{year}{1972}).

\bibitem[{\citenamefont{Gring et~al.}(2012)\citenamefont{Gring, Kuhnert,
  Langen, Kitagawa, Rauer, Schreitl, Mazets, Smith, Demler, and
  Schmiedmayer}}]{GringKuhnert}
\bibinfo{author}{\bibfnamefont{M.} \bibnamefont{Gring}},
  \bibinfo{author}{\bibfnamefont{M.} \bibnamefont{Kuhnert}},
  \bibinfo{author}{\bibfnamefont{T.} \bibnamefont{Langen}},
  \bibinfo{author}{\bibfnamefont{T.} \bibnamefont{Kitagawa}},
  \bibinfo{author}{\bibfnamefont{B.} \bibnamefont{Rauer}},
  \bibinfo{author}{\bibfnamefont{M.} \bibnamefont{Schreitl}},
  \bibinfo{author}{\bibfnamefont{I.} \bibnamefont{Mazets}},
  \bibinfo{author}{\bibfnamefont{D.~A.} \bibnamefont{Smith}},
  \bibinfo{author}{\bibfnamefont{E.} \bibnamefont{Demler}}, \bibnamefont{and}
  \bibinfo{author}{\bibfnamefont{J.} \bibnamefont{Schmiedmayer}},
  \bibinfo{journal}{Science} \textbf{\bibinfo{volume}{337}},
  \bibinfo{pages}{1318} (\bibinfo{year}{2012}).

\end{thebibliography}
\end{document}